\documentstyle[aps,12pt]{revtex}
\begin{document}
\title{Neutron density distributions for
atomic parity nonconservation experiments}
\bigskip
\author{D. Vretenar$^{1,2}$, G.A. Lalazissis$^{1,3}$, 
and P. Ring$^{1}$}
\bigskip
\address{
$^{1}$ Physik-Department der Technischen Universit\"at M\"unchen,
D-85748 Garching, Germany\\
$^{2}$ Physics Department, Faculty of Science, University of
Zagreb, 10 000 Zagreb, Croatia\\
$^{3}$ Physics Department, Aristotle University of Thessaloniki,
Thessaloniki GR-54006, Greece\\}
\maketitle
\bigskip
\bigskip
\begin{abstract}
The neutron distributions of Cs, Ba, Yb and Pb isotopes
are described in the framework of relativistic mean-field theory.
The self-consistent ground state proton and neutron 
density distributions are calculated with the relativistic 
Hartree-Bogoliubov model. The binding energies, the proton and neutron 
radii, and the quadrupole deformations are compared with available 
experimental data, as well as with recent theoretical studies of 
the nuclear structure corrections to the weak charge in atomic 
parity nonconservation experiments.
\end{abstract}
\vspace{1 cm}
{PACS number(s):} {32.80.Ys, 12.15.Ji, 21.10.Gv, 21.60.-n}\\
\newpage       
\section{Introduction}

Measurements of parity nonconservation (PNC) effects in 
intermediate and heavy atomic systems might provide very stringent
tests of the Standard model of electroweak interactions. 
Nuclei and electrons are bound into atoms by electromagnetic 
interactions that do not violate parity. Parity nonconservation
arises from the interference of the $\gamma$ and $Z^0$ gauge 
boson exchange between the nucleus and the atomic electrons. 
The dominant part of the PNC interaction results from the 
coupling of the axial electronic current to the vector nuclear 
current, and it allows normally forbidden transitions. The 
nuclear vector current is conserved and the PNC effects 
provide a measure of the electroweak coupling of the 
elementary quarks. An extensive analysis of the sensitivity
of a variety of low-energy PNC observables to possible new physics 
phenomena has been recently reported in Ref.~\cite{Mus.99}.

The nucleus acts as a source of the weak charge $Q_W$. The 
quantity which is measured in experiments is the electric 
dipole amplitude between two electronic states which, in
the absence of the PNC electron-nucleus interaction, 
would have the same parity. This observable can be 
parameterized~\cite{Mus.99,PW.99}
\begin{equation}
\epsilon_{PNC}~=~\xi Q_W~=~\xi\left [ Q^{St.Mod}_W + 
\Delta Q^{n-p}_W \right ],
\label{xi}
\end{equation}
where $\xi$ is a coefficient that depends on atomic structure,
$Q^{St.Mod}_W$ is the Standard model prediction for the weak
charge, and $\Delta Q^{n-p}_W$ results from the difference
between the neutron and proton density distributions in the
nucleus. At tree level 
\begin{equation}
Q^{St.Mod}_W = (1-4\sin^2\theta_W)Z - N.
\end{equation}
Z and N are the numbers of protons and neutrons, respectively, 
and $\theta_W$ is the weak mixing angle. The finite size of 
the nucleus modifies $Z\rightarrow q_{{p}} Z$ and 
$N\rightarrow q_{{n}} N$,
where
\begin{equation}
q_{{n} ({p})} = \int f(r)
\rho_{{n} ({p})}(r) d^{3} r .
\label{weakcharge}
\end{equation}
The function $f(r)$ describes the spatial variation of the 
electron axial transition matrix element over the nucleus.
The difference between the neutron $\rho_n$ and proton 
$\rho_p$ density distributions modifies the effective 
weak charge 
\begin{equation}
\Delta Q^{n-p}_W =  N(1-q_{{n}}/q_{{p}}).
\label{deltaQ}
\end{equation}
The two principal factors, therefore, which limit the precision 
with which the weak charge $Q_W$ can be determined from atomic 
experiments are: (i) the accuracy of atomic theory calculations
($\xi$ in Eq.~(\ref{xi}) and the function 
$f(r)$ in Eq.~(\ref{weakcharge})), and (ii) the nuclear structure
effects, i.e. the uncertainties in the neutron and proton density
distributions. At present, the level of accuracy of atomic structure
calculations constitutes the predominant uncertainty in the 
interpretation of atomic PNC observables~\cite{Mus.99}. For example,
the most accurate experimental result for the weak charge of the
Cs nucleus is $Q^{expt}_W = - 72.06 (28)_{expt} 
(34)_{theory}$~\cite{BW.99}.
The Standard model value including radiative corrections is
$Q^{St.Mod}_W = - 73.20 (13)$. It has been suggested~\cite{Dzuba}
that the measurement of PNC effects along an isotope chain
might represent a useful method to 
circumvent the atomic theory uncertainty.
The ratios of the electric dipole amplitudes for different isotopes
should not depend on the details of atomic structure. In addition
to the obvious N dependence, these ratios will be sensitive 
to the value of the Weinberg angle and, eventually, to additional 
parameters of a more general electroweak theory~\cite{Mus.99,For.90}.
The shortcoming of this method, however, is that the ratios \
of PNC observables 
will also be sensitive to changes in the neutron distribution $\rho_n$
along the isotope chain. The PNC interaction is directly proportional
to the overlap between electrons and neutrons.
And while extremely accurate data on 
proton distributions in nuclei are obtained from
electron scattering experiments, data of comparable precision on 
neutron density distributions are presently not available. 
It is much more difficult to measure the distribution of neutrons,
though more recently accurate data on differences 
in radii of the neutron and proton density distributions have been
obtained~\cite{Suz.95,Kra.99}.

A number of studies have been reported on the nuclear structure
effects in the interpretation of atomic PNC experiments in 
Cs and Pb isotope chains. The basic theoretical framework has
been analyzed in Ref.~\cite{For.90}, and it was pointed out that
the uncertainties in the neutron distributions would have
to be significantly reduced, if isotope ratios are to 
be used in high-precision evaluation of the atomic
PNC parameters. On the other hand, it was also emphasized
that atomic PNC experiments might provide a model independent
tool for studying variations in neutron distributions.
In a more extensive analysis of electroweak parameters and
nuclear structure effects~\cite{Pol.92}, a number of 
relativistic and nonrelativistic nuclear model calculations
were performed for the chain of Pb isotopes. The conclusion 
was that the spread among model predictions for the 
sizes of nuclear structure effects may preclude using Pb 
isotope ratios to test the Standard model at better than 
a 1\% level. The theoretical analysis included the 
Hartree-Fock-Bogoliubov model with Gogny forces,
various Hartree-Fock Skyrme effective interactions, 
as well as the relativistic mean-field model.
The Hartree-Fock Skyrme model, with the inclusion of 
deformation effects, was used in Ref.~\cite{Chen} 
to calculate the proton and neutron densities in 
the isotope chain $^{125}$Cs - $^{139}$Cs. The good
agreement with the experimental charge radii, binding 
energies, and ground-state spins, led to the estimate
that the uncertainties in the differences of the neutron radii
may cause uncertainties in the ratio of weak charges of 
less than $10^{-3}$, safely smaller than the anticipated
experimental error. In a detailed analysis of Ref.~\cite{Mille}
it was shown that extremely high-precision measurements of 
PNC effects in atomic Yb appear possible, and it 
was suggested that a comparison of PNC observables in the wide 
range of stable isotopes of Yb may provide a unique test 
of the Standard model. The recent experimental result
for the weak charge of Cs~\cite{BW.99} has prompted the
analysis of the effects of neutron spatial distribution
on atomic PNC in Cs~\cite{PW.99}. Several theoretical
models have been used to calculate the modifications to 
the nuclear weak charge due to small differences between
the spatial distributions of neutrons and protons in the
Cs nucleus. It has been shown that the nuclear structure 
uncertainties are smaller than those arising from atomic
theory calculations, but are comparable to the present 
uncertainties in the Standard model prediction. The 
sensitivity of atomic PNC to nuclear structure effects
in single isotopes and isotope chains has also been 
extensively discussed in Refs.~\cite{Mus.99} and~\cite{Hor.99}.

In the comparison of the theoretical models that have been used
in the analysis of nuclear structure effects, it has been pointed
out~\cite{PW.99,Pol.92} that relativistic mean field models typically
generate significantly larger neutron radii, and thus would predict
much larger corrections to the nuclear weak charge. Models of 
nuclear structure based on relativistic hadrodynamics~\cite{SW.97}
have attracted much interest in the last decade. In particular, 
the relativistic mean-field model has been very successfully applied
in the description of a large variety of nuclear structure
phenomena in spherical and deformed nuclei~\cite{Rin.96}. 
The model has more recently been extended to include both 
mean-field and pairing correlations in a unified framework: the
relativistic Hartree-Bogoliubov model~\cite{MR.96,PVL.97}. A unified
and self-consistent description of the long range mean-field
correlations and the pairing field is especially important for
the structure of open shell nuclei further away from the 
line of $\beta$-stability. In addition, in the last few years
several new and more accurate parameter sets of meson masses and
meson-nucleon coupling constants for the effective mean-field
Lagrangian have been derived. These new effective forces have 
been used not only for a more accurate description of 
properties of $\beta$-stable nuclei, but they have also been
applied in studies of isovector properties of exotic nuclei
with large isospin values, of the structure of superheavy 
nuclei, and properties of excited states in the relativistic
random-phase approximation and the time-dependent relativistic
mean-field model. 

In a recent study~\cite{Vre.99} we have applied the 
relativistic Hartree-Bogoliubov model to the description
of parity-violating elastic electron scattering (PVES)
on neutron-rich Ne, Na, Ni and Sn isotopes. For the elastic
scattering of 850 MeV electrons on these nuclei, the
parity-violating asymmetry parameters have been calculated
using a relativistic optical model with inclusion of
Coulomb distortion effects. The asymmetry parameters
for chains of isotopes have been compared, and their relation
to the Fourier transforms of neutron densities has been
studied. The important impact that PVES measurement of 
neutron densities could have on atomic PNC experiments
has been recently discussed in Ref.~\cite{Hor.99}. 
Since the sensitivity of the neutron distribution
appears to be approximately the same for PVES and atomic
PNC, it would of course be ideal if density 
distributions for PNC experiments were determined from PVES. 
In the present work we use the relativistic Hartree-Bogoliubov 
model to calculate ground-state proton and neutron distributions
for the isotopic chains that have been suggested for atomic
PNC experiments: Cs, Ba, Yb and Pb. Calculated quantities
will be compared with available experimental data, as well
as with previous analyses of the nuclear structure corrections
to the weak charge~\cite{PW.99,Pol.92,Chen}. 
In Section II we present an outline of the relativistic 
Hartree-Bogoliubov model. The neutron densities for atomic
PNC experiments are calculated and analyzed in Section III.

\section{The Relativistic Hartree-Bogoliubov model}

The relativistic mean field theory is based on simple concepts:
nucleons are described as point particles, the theory is fully Lorentz
invariant, the nucleons move independently in
mean fields which originate from the nucleon-nucleon interaction.
Conditions of causality and Lorentz invariance impose that the
interaction is mediated by the
exchange of point-like effective mesons, which couple to the nucleons
at local vertices. The single-nucleon dynamics is described by the
Dirac equation
\begin{equation}
\label{statDirac}
\left\{-i\mbox{\boldmath $\alpha$}
\cdot\mbox{\boldmath $\nabla$}
+\beta(m+g_\sigma \sigma)
+g_\omega \omega^0+g_\rho\tau_3\rho^0_3
+e\frac{(1-\tau_3)}{2} A^0\right\}\psi_i=
\varepsilon_i\psi_i.
\end{equation}
$\sigma$, $\omega$, and
$\rho$ are the meson fields, and $A$ denotes the electromagnetic potential.
$g_\sigma$ $g_\omega$, and $g_\rho$ are the corresponding coupling
constants for the mesons to the nucleon.
The lowest order of the quantum field theory is the {\it
mean-field} approximation: the meson field operators are
replaced by their expectation values. The sources
of the meson fields are defined by the nucleon densities
and currents.  The ground state of a nucleus is described
by the stationary self-consistent solution of the coupled
system of the Dirac~(\ref{statDirac})and Klein-Gordon equations:
\begin{eqnarray}
\left[ -\Delta +m_{\sigma }^{2}\right] \,\sigma ({\bf r}) &=&-g_{\sigma
}\,\rho _{s}({\bf r})-g_{2}\,\sigma ^{2}({\bf r})-g_{3}\,\sigma ^{3}({\bf r})
\label{messig} \\
\left[ -\Delta +m_{\omega }^{2}\right] \,\omega ^{0}({\bf r}) &=&g_{\omega
}\,\rho _{v}({\bf r})  \label{mesome} \\
\left[ -\Delta +m_{\rho }^{2}\right] \,\rho ^{0}({\bf r}) &=&g_{\rho }\,\rho
_{3}({\bf r})  \label{mesrho} \\
-\Delta \,A^{0}({\bf r}) &=&e\,\rho _{p}({\bf r}),  \label{photon}
\end{eqnarray}
for the sigma meson, omega meson, rho meson and photon field, respectively.
Due to charge conservation, only the 3rd-component of the isovector rho
meson contributes. The source terms in equations (\ref{messig}) to (\ref
{photon}) are sums of bilinear products of baryon amplitudes, and they 
are calculated in the {\it no-sea} approximation, i.e. the Dirac sea
of negative energy states does not contribute to the nucleon densities
and currents. Due to time reversal invariance,
there are no currents in the static solution for an even-even
system, and therefore the spatial
vector components \mbox{\boldmath $\omega,~\rho_3$} and
${\bf  A}$ of the vector meson fields vanish.
The quartic potential 
\begin{equation}
U(\sigma )~=~\frac{1}{2}m_{\sigma }^{2}\sigma ^{2}+\frac{1}{3}g_{2}\sigma
^{3}+\frac{1}{4}g_{3}\sigma ^{4}  \label{usigma}
\end{equation}
introduces an effective density dependence. The non-linear
self-interaction of the $\sigma$ field is essential for 
a quantitative description of properties of finite nuclei.

In addition to the self-consistent mean-field
potential, pairing correlations have to be included in order to
describe ground-state properties of open-shell nuclei.
For nuclei close to the $\beta$-stability
line, pairing has been included in the relativistic
mean-field model in the form of a simple BCS
approximation~\cite{GRT.90}. For more exotic nuclei further
away from the stability line, however, the BCS model presents 
only a poor approximation. In particular, in order to 
correctly reproduce density distributions in neutron-rich 
nuclei, mean-field and pairing correlations have to be 
described in a unified framework: the Hartree-Fock-Bogoliubov
model or the relativistic Hartree-Bogoliubov (RHB) model.
In the unified framework 
the ground state of a nucleus $\vert \Phi >$ is represented
by the product of independent single-quasiparticle states.
These states are eigenvectors of the
generalized single-nucleon Hamiltonian which
contains two average potentials: the self-consistent mean-field
$\hat\Gamma$ which encloses all the long range particle-hole ({\it ph})
correlations, and a pairing field $\hat\Delta$ which sums
up the particle-particle ({\it pp}) correlations.
In the Hartree approximation for
the self-consistent mean field, the relativistic
Hartree-Bogoliubov equations read
\bigskip
\begin{eqnarray}
\label{equ.2.2}
\left( \matrix{ \hat h_D -m- \lambda & \hat\Delta \cr
		-\hat\Delta^* & -\hat h_D + m +\lambda} \right)
		\left( \matrix{ U_k({\bf r}) \cr V_k({\bf r}) } \right) =
		E_k\left( \matrix{ U_k({\bf r}) \cr V_k({\bf r}) } \right).
\end{eqnarray}
\bigskip
where $\hat h_D$ is the single-nucleon Dirac
Hamiltonian (\ref{statDirac}), and $m$ is the nucleon mass.
The chemical potential $\lambda$  has to be determined by
the particle number subsidiary condition in order that the
expectation value of the particle number operator
in the ground state equals the number of nucleons. The column
vectors denote the quasi-particle spinors and $E_k$
are the quasi-particle energies.
The pairing field $\hat\Delta $ in (\ref{equ.2.2}) is defined
\begin{equation}
\label{equ.2.5}
\Delta_{ab} ({\bf r}, {\bf r}') = {1\over 2}\sum\limits_{c,d}
V_{abcd}({\bf r},{\bf r}') \sum_{E_k>0} U_{ck}^*({\bf r})V_{dk}({\bf r}'),
\end{equation}
where $a,b,c,d$ denote quantum numbers
that specify the Dirac indices of the spinors,
$V_{abcd}({\bf r},{\bf r}')$ are matrix elements of a
general two-body pairing interaction.
The RHB equations are solved self-consistently, with
potentials determined in the mean-field approximation from
solutions of Klein-Gordon equations for the meson fields.
The current version of the model~\cite{LVR.99}
describes axially symmetric deformed shapes.
The Dirac-Hartree-Bogoliubov equations and the equations for the
meson fields are solved by expanding the nucleon spinors
$U_k({\bf r})$ and $V_k({\bf r})$,
and the meson fields in terms of the eigenfunctions of a
deformed axially symmetric oscillator potential~\cite{GRT.90}.
The calculations for the present analysis have been performed
by an expansion in 12 oscillator shells for the fermion fields,
and 20 shells for the boson fields.
A simple blocking procedure is used in the calculation of
odd-proton and/or odd-neutron systems. The blocking calculations
are performed without breaking the time-reversal symmetry.

The input parameters of the RHB model are the coupling constants and the
masses for the effective mean-field Lagrangian, and the effective
interaction in the pairing channel. In most applications we have
used  the NL3 effective interaction \cite{LKR.97} for the RMF
Lagrangian. Properties calculated with NL3 indicate that this is probably
the best effective interaction so far, both for nuclei at and away from the
line of $\beta $-stability. For the pairing field we employ the
pairing part of the Gogny interaction
\begin{equation}
V^{pp}(1,2)~=~\sum_{i=1,2}
e^{-(( {\bf r}_1- {\bf r}_2)
/ {\mu_i} )^2}\,
(W_i~+~B_i P^\sigma
-H_i P^\tau -
M_i P^\sigma P^\tau),
\end{equation}
with the set D1S \cite{BGG.84} for the parameters
$\mu_i$, $W_i$, $B_i$, $H_i$ and $M_i$ $(i=1,2)$.
This force has been very carefully adjusted to the pairing
properties of finite nuclei all over the periodic table.
In particular, the basic advantage of the Gogny force
is the finite range, which automatically guarantees a proper
cut-off in momentum space.
The RHB model with the NL3+D1S effective interaction 
has been applied in studies of the
neutron halo phenomenon in light nuclei \cite{PVL.97},
properties of light nuclei near the neutron-drip line \cite{LVP.98},
ground state properties of Ni and Sn isotopes \cite{LVR.98},
the deformation and shape coexistence phenomena that result from the
suppression of the spherical N=28 shell gap in neutron-rich nuclei
\cite{LVR.98a}, the structure of proton-rich nuclei and the
phenomenon of ground state proton emission \cite{LVR.99,VLR.99,LVR.99a}.
In particular, it has been shown that neutron radii, calculated
with the RHB NL3+D1S model, are in excellent agreement with experimental
data \cite{LVP.98,LVR.98}.

\section{Neutron densities for atomic PNC experiments}

In this section we apply the relativistic Hartree-Bogoliubov model,
with the NL3+D1S effective interaction, in the calculation of
ground-state proton and neutron densities of Cs, Ba, Yb and Pb isotopes.
These elements have been considered for measurements of 
isotope ratios of observables in atomic PNC 
experiments~\cite{Mus.99,For.90,Pol.92,Chen,Mille}.

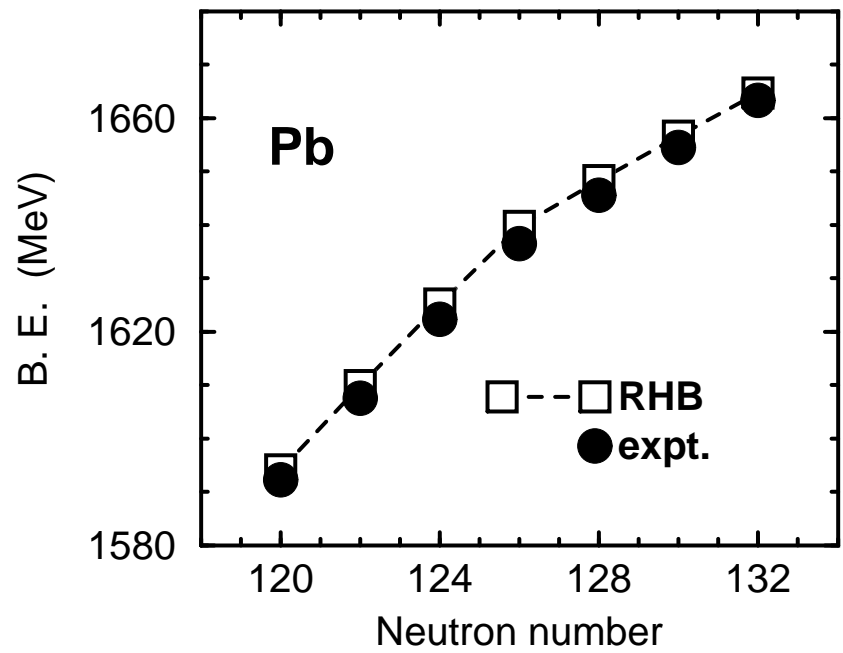
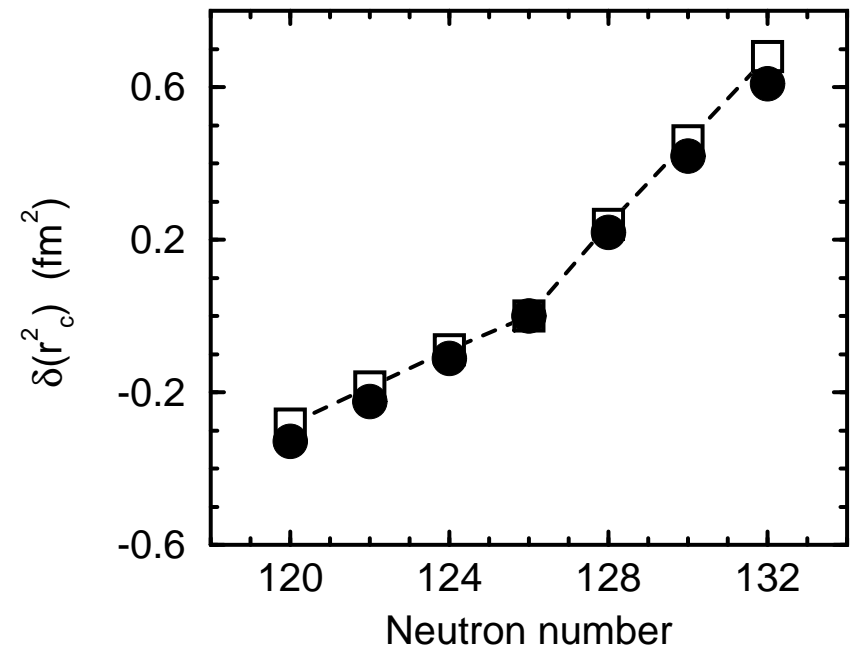
\begin{figure}
\caption{Binding energies (left), and isotope shifts
of the charge radii (right) of Pb nuclei, calculated 
with the NL3 + Gogny D1S effective interaction.
The theoretical values are compared with the experimental
binding energies~{\protect \cite{AW.95}} and charge radii~{\protect 
\cite{Ott.89}}.}
\label{figA}
\end{figure}

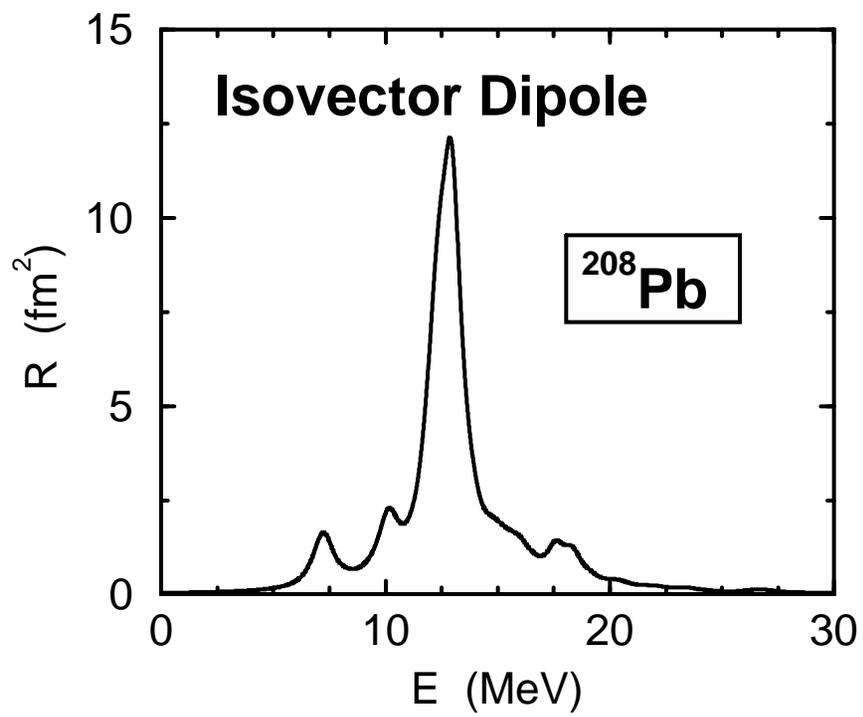
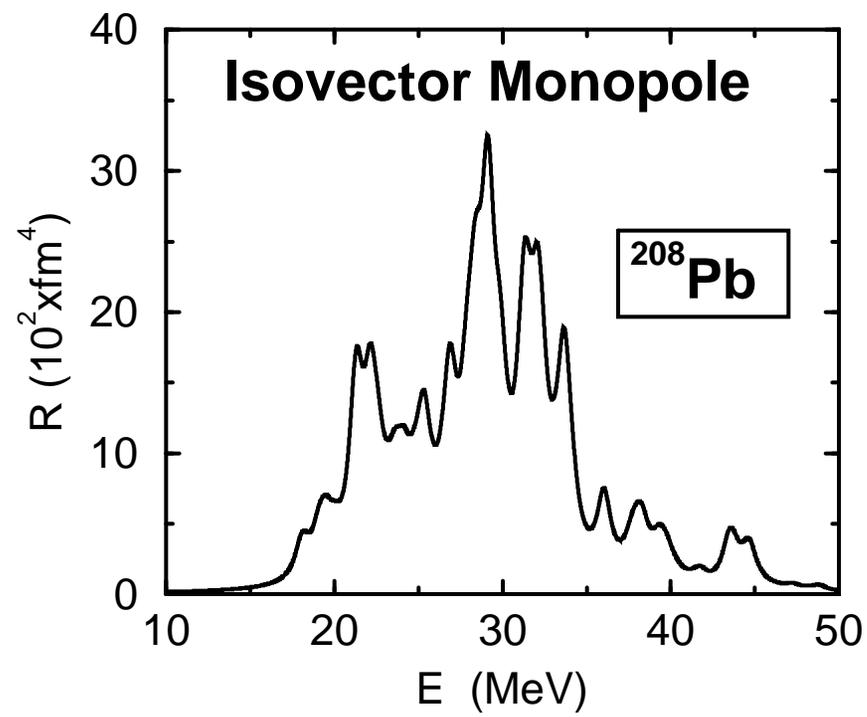
\begin{figure}
\caption{The isovector dipole and monopole strength distributions
in $^{208}$Pb, calculated with the NL3 effective
interaction in the relativistic RPA framework.}
\label{figB}
\end{figure}

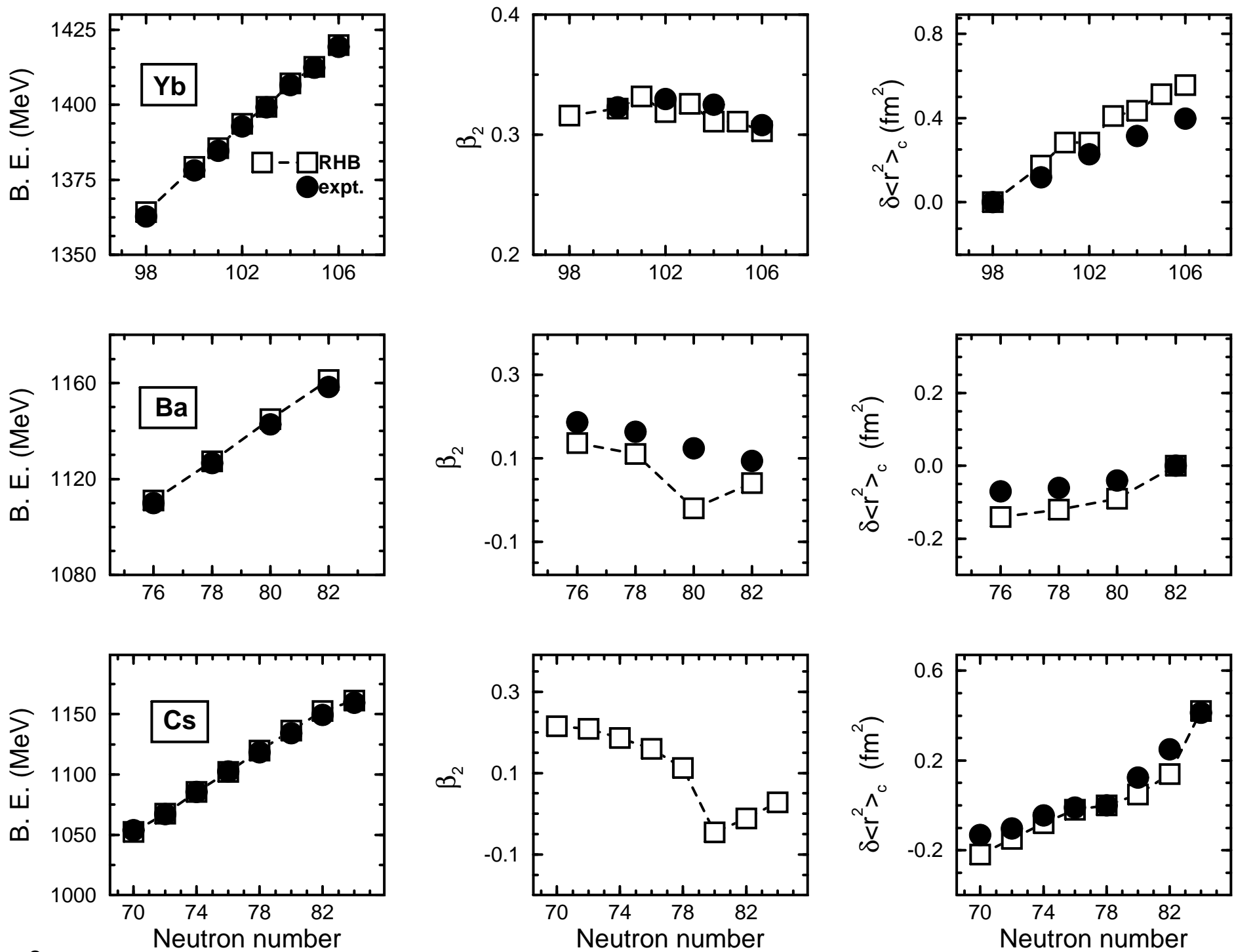
\begin{figure}
\caption{Binding energies (left), quadrupole deformations
(center), and isotope shifts of the charge radii (right),
for the Yb, Ba and Cs isotope chains. The results of 
the self-consistent RHB calculations are compared with 
the available experimental data~{\protect \cite{AW.95,Ram.87,Ott.89}.}}
\label{figC}
\end{figure}

Due to the large uncertainties in the experimental data on
ground-state isovector properties, most of the effective
interactions used in the non-relativistic Hartre-Fock model, and
the effective Lagrangians
of the relativistic mean-field model, have not been specifically
designed to describe neutron density distributions. Only more
recently have isovector properties been included in the set 
of data on which the effective interactions are adjusted. 
In the relativistic mean-field model, perhaps the most 
accurate set of meson masses and meson nucleon coupling 
constants is NL3~\cite{LKR.97}. In Ref.~\cite{LRR.98}
this effective force has been used in the relativistic
mean-field + BCS model to calculate the ground-state 
properties of 1315 even-even nuclei with $10\leq Z\leq 98$.
The calculated quantities include the total binding
energies, rms radii, quadrupole and hexadecupole deformations.
In order to illustrate the quality of the results that are 
obtained with the NL3 interaction, in Fig.~\ref{figA} we 
display the binding energies and isotope shifts 
of the charge radii of Pb nuclei. The calculation has been 
performed with the RHB model, and the theoretical values 
are compared with the experimental data on binding 
energies~\cite{AW.95} and charge radii~\cite{Ott.89}.
The agreement between theory and experiment is very good,
and we notice in particular that the calculated radii
reproduce the anomalous kink in the isotopic shifts at
$^{208}$Pb~\cite{Ott.89,SLR.93}. The NL3 effective 
force has also been used in studies of properties 
of excited states. By using the time-dependent
relativistic mean-field model~\cite{Vre.97}, and
the relativistic random-phase approximation~\cite{Vre.00},
it has been shown that a self-consistent calculation 
with the NL3 interaction reproduces the excitation 
energies of giant resonances in spherical nuclei. 
Particularly relevant for our present study are
isovector giant resonances. In Fig.~\ref{figB} we show
the isovector dipole and monopole strength distributions
in $^{208}$Pb, calculated with the NL3 effective
interaction in the relativistic RPA framework. The calculated
peak energy for the dipole resonance $E_p = 12.9$ MeV has to 
be compared with the experimental value of the excitation 
energy $13.3\pm 0.1$ MeV~\cite{Rit.93}. The calculated 
isovector monopole strength is much more fragmented, 
but the centroid is also in good agreement 
with the experimental value for the IV GMR:
$26\pm 3$ MeV~\cite{GBM.90}. In Fig~\ref{figC} we plot
the binding energies, quadrupole deformations, 
and isotope shifts of the charge radii 
for the Yb, Ba and Cs isotope chains. The results of
the self-consistent RHB NL3+D1S calculations are compared with
the available experimental data~\cite{AW.95,Ott.89,Ram.87}.
The overall agreement between the theoretical predictions and
experimental data is excellent. 
The results shown in Figs.~\ref{figA}-~\ref{figC} demonstrate that
the RHB model with the NL3+D1S effective interaction is adequate
for the calculation of ground-state neutron densities. 

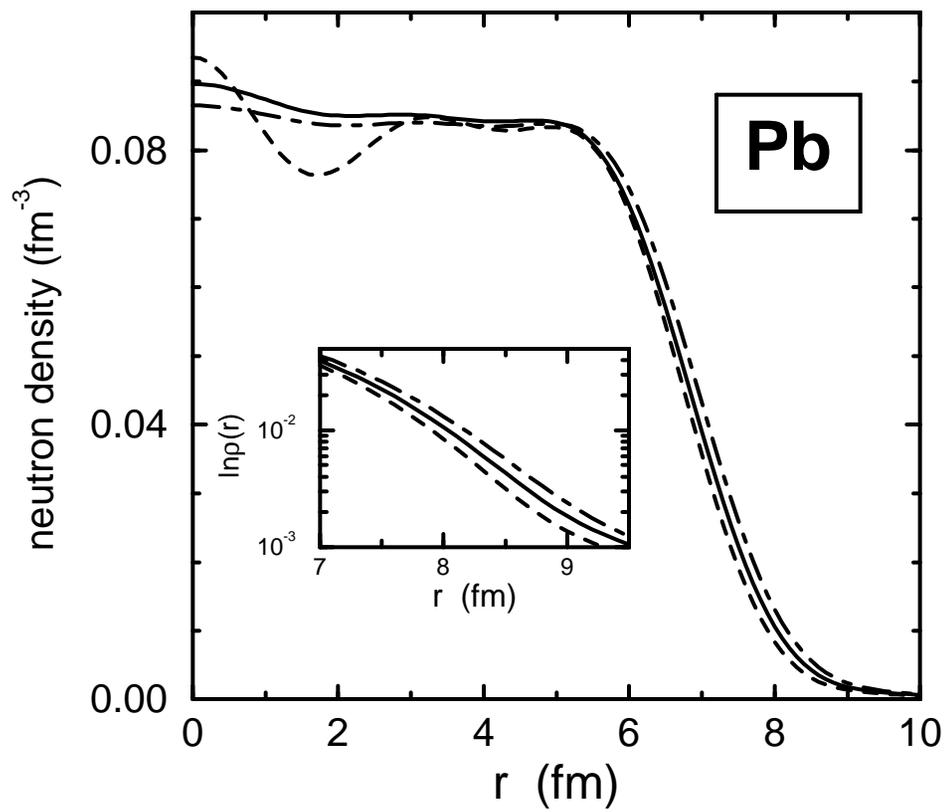
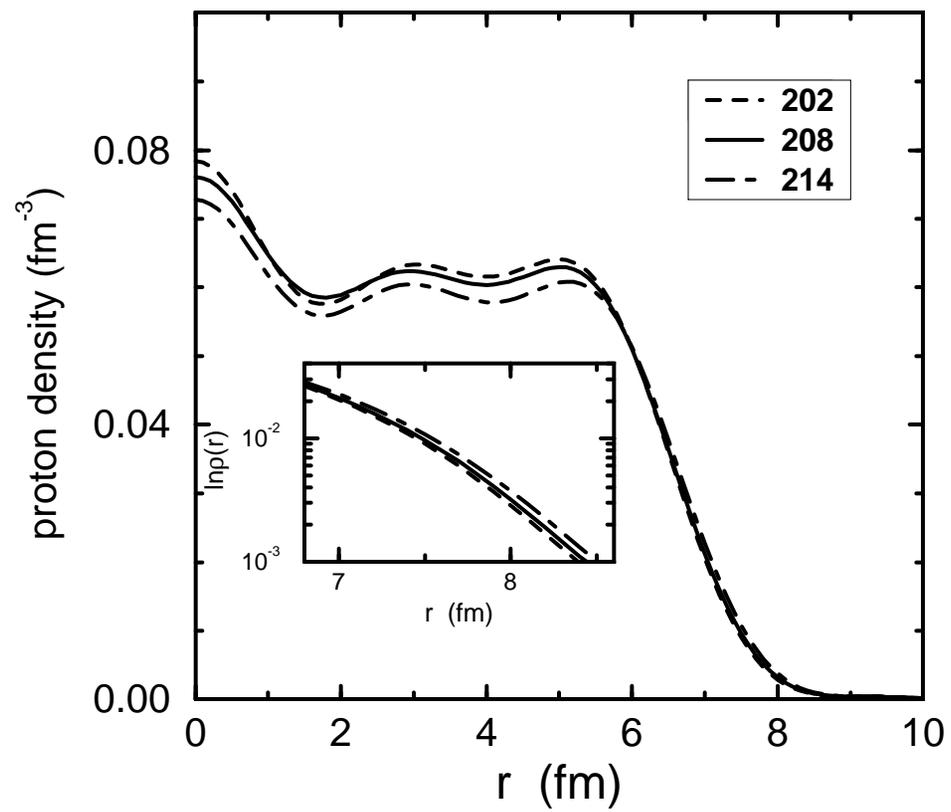
\begin{figure}
\caption{Self-consistent RHB neutron and proton 
ground-state density distributions for $^{202,208,214}$Pb.
In the inserts the same densities in the surface region 
are plotted on the logarithmic scale.}
\label{figD}
\end{figure}

The self-consistent RHB neutron and proton 
ground-state density distributions for the isotopes 
$^{202,208,214}$Pb are shown in Fig.~\ref{figD}. We will 
eventually compare the calculated rms radii for $^{202-214}$Pb, 
but here we notice the trend:
the neutron radii increase  with 
neutron number, while the neutron densities in the bulk
region do not change much, except for some shell effects in
the central region. The proton densities on the other hand,
decrease in the bulk region as more neutrons are added 
on the surface. This effect, of course, has its origin 
in the symmetry energy. Therefore, one also observes
an increase of proton radii, but much less pronounced 
than in the case of neutron densities. The difference
in the trend of neutron and proton radii is emphasized
in the inserts of Fig.~\ref{figD}, where the densities
in the surface region are plotted on the logarithmic scale.

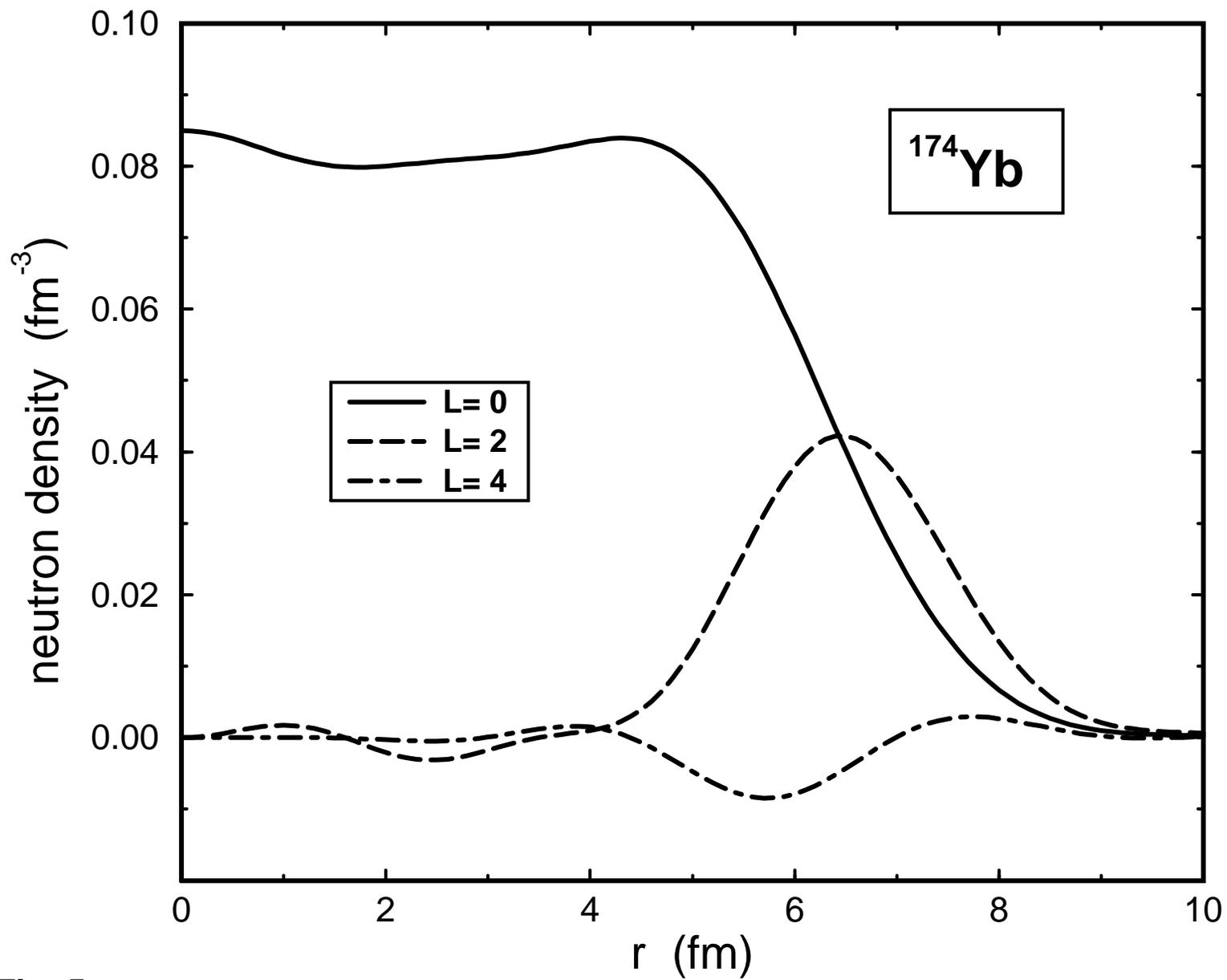
\begin{figure}
\caption{The monopole, quadrupole and hexadecupole components
of the self-consistent RHB neutron ground-state density distribution
in the nucleus $^{174}$Yb.}
\label{figE}
\end{figure}

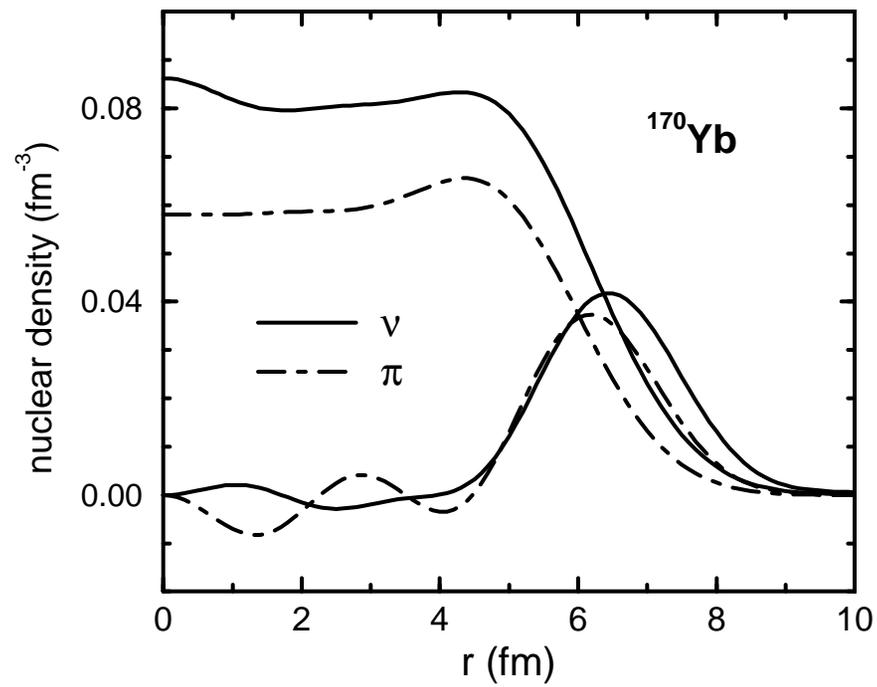
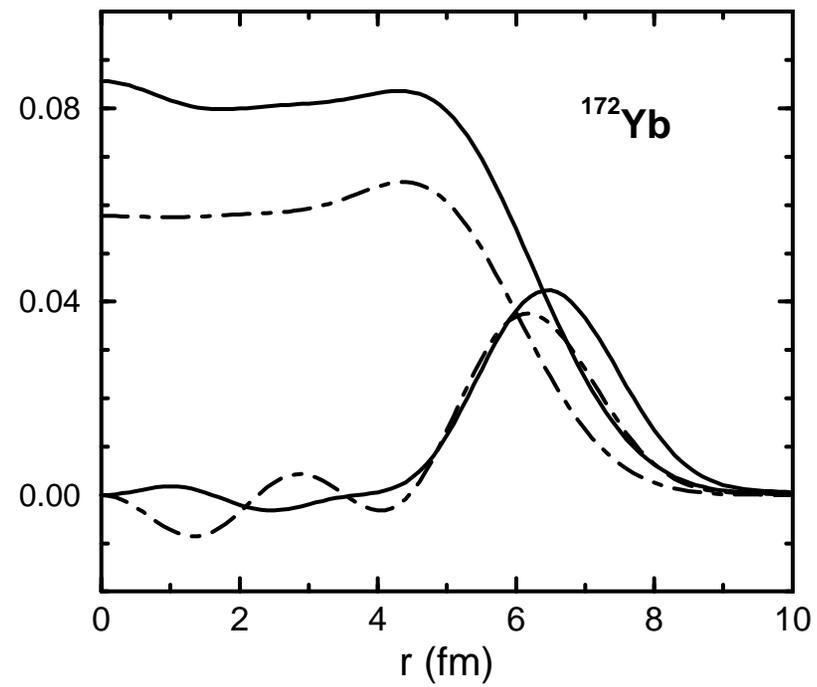
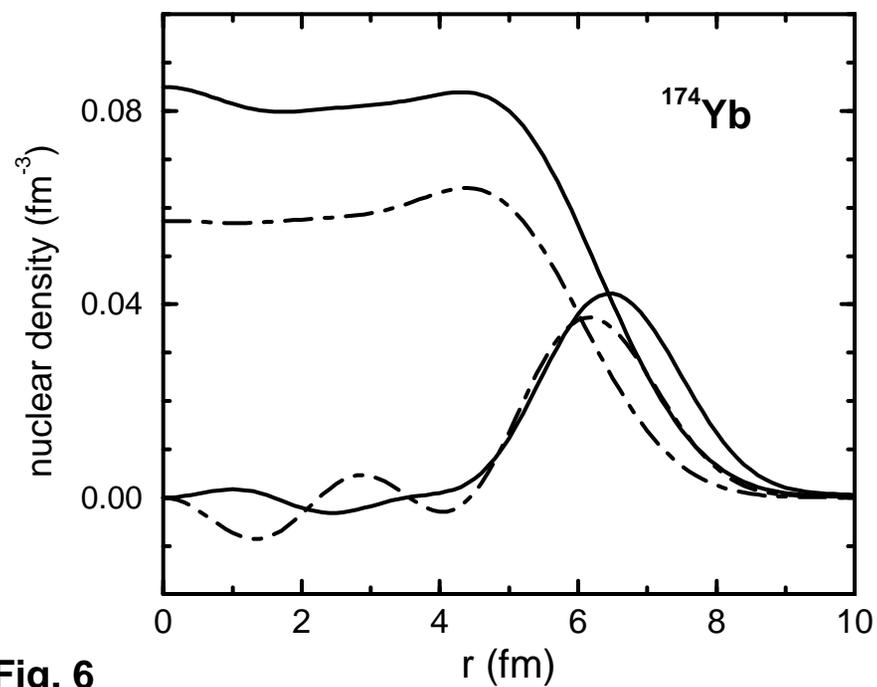
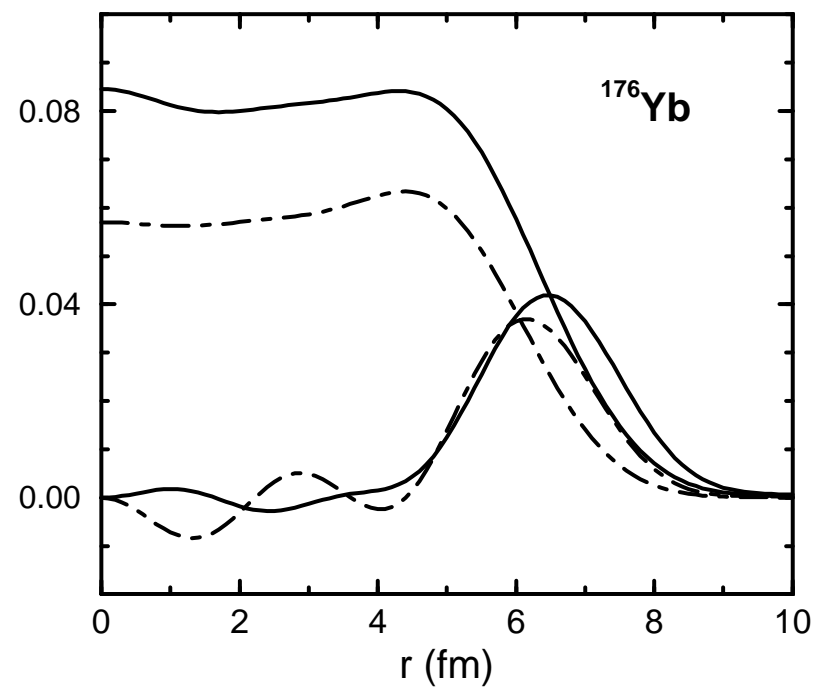
\begin{figure}
\caption{Monopole and quadrupole components of the 
self-consistent RHB neutron (solid line) and proton (dot-dashed line)
ground-state density distributions for the even-N Yb isotopes.}
\label{figF}
\end{figure}

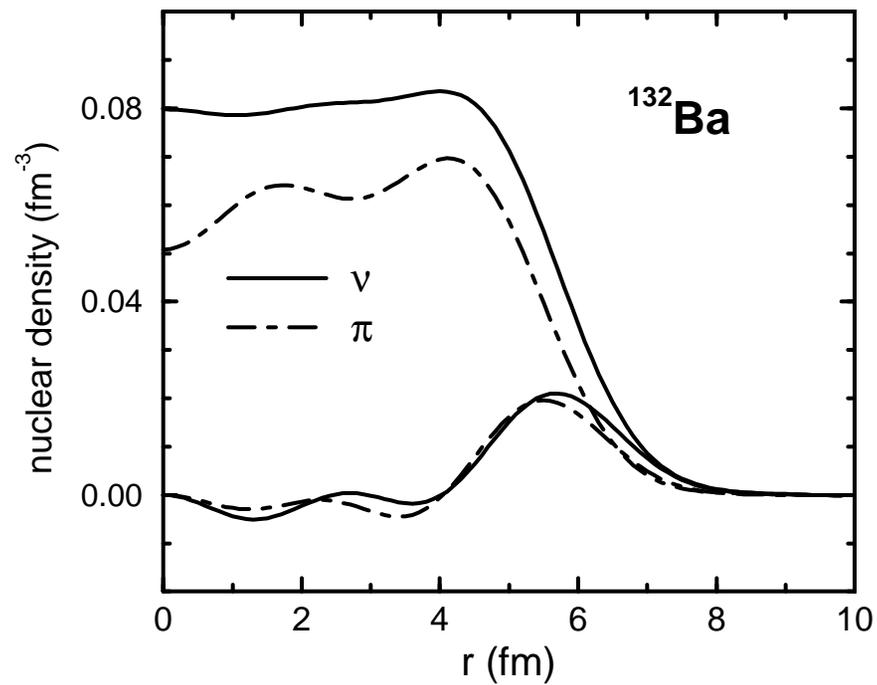
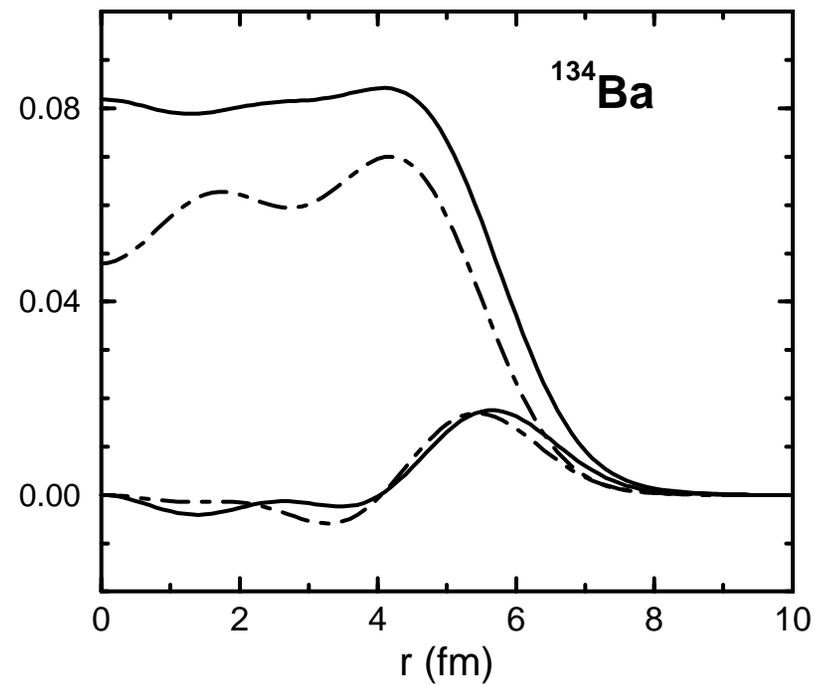
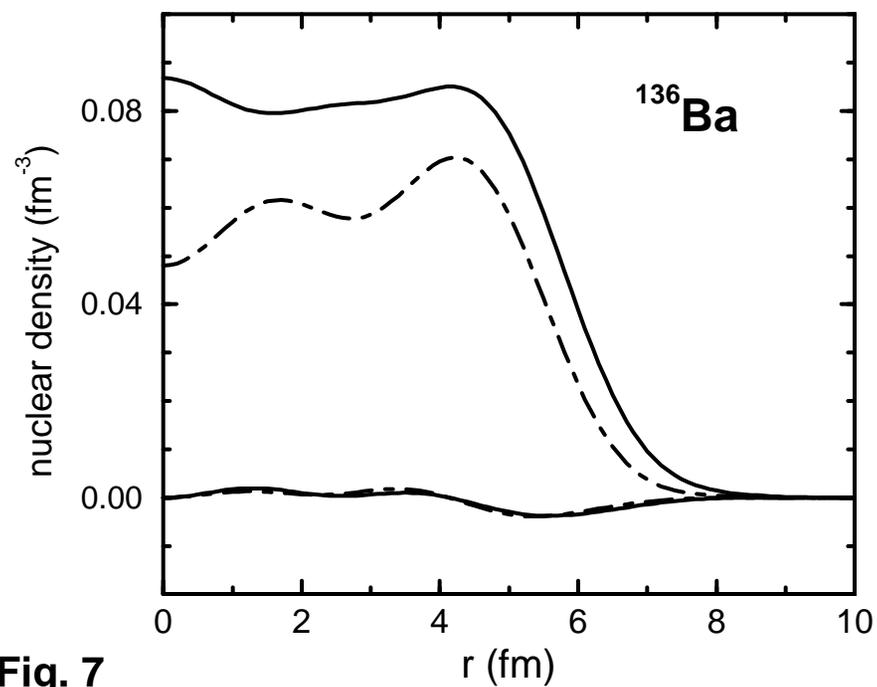
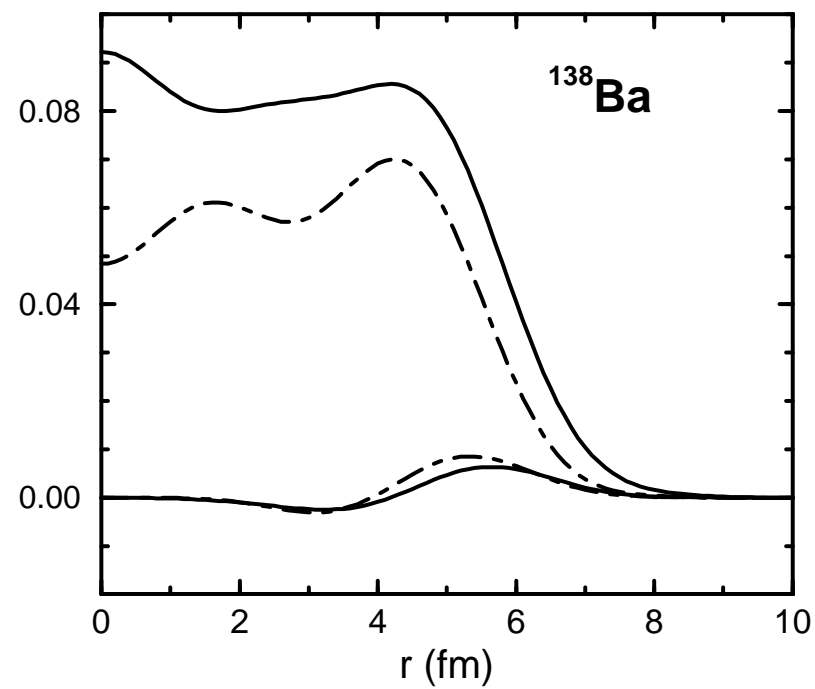
\begin{figure}
\caption{Same as in Fig. \ref{figF}, but for the Ba isotopes.}
\label{figG}
\end{figure}

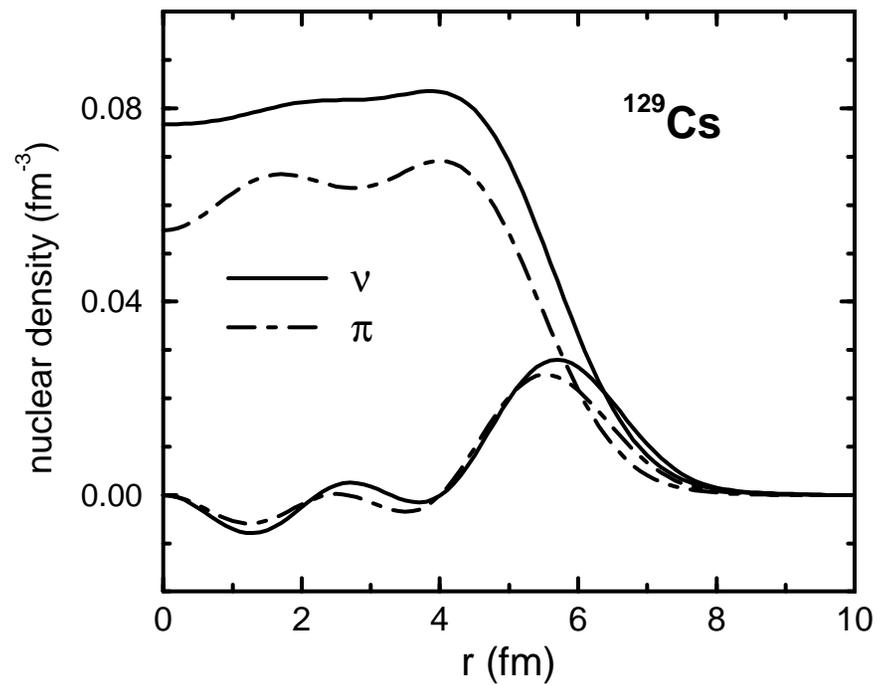
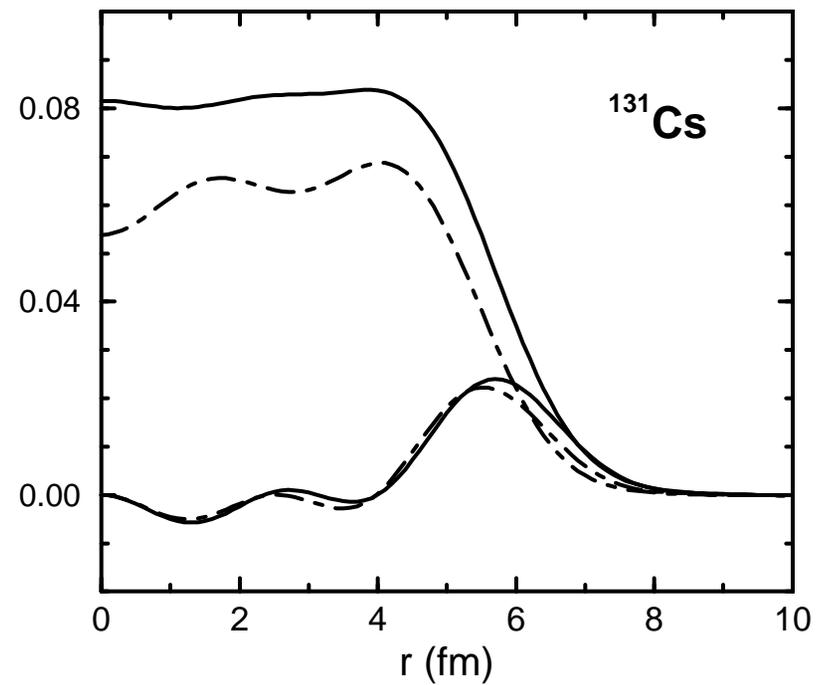
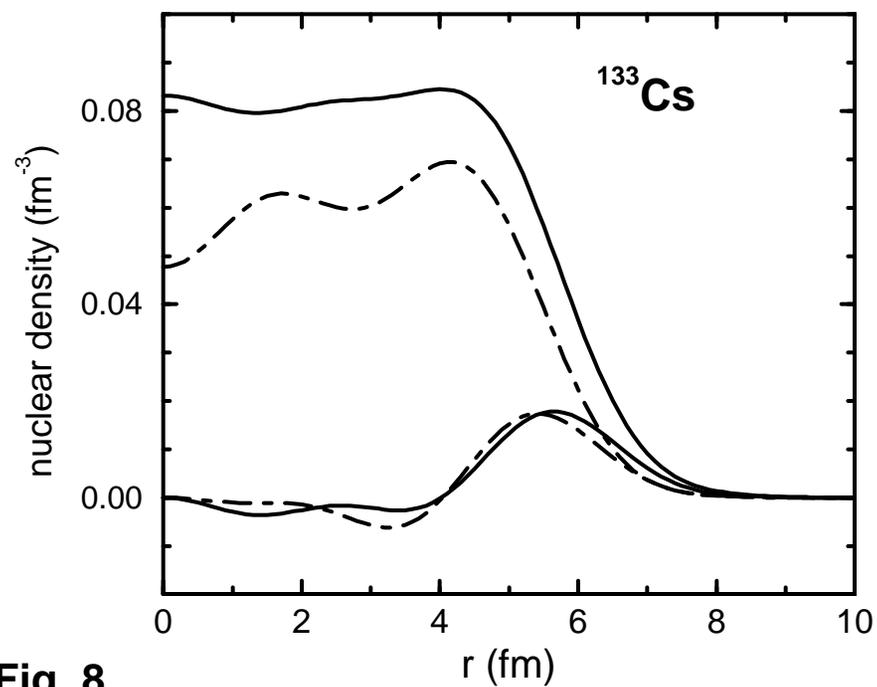
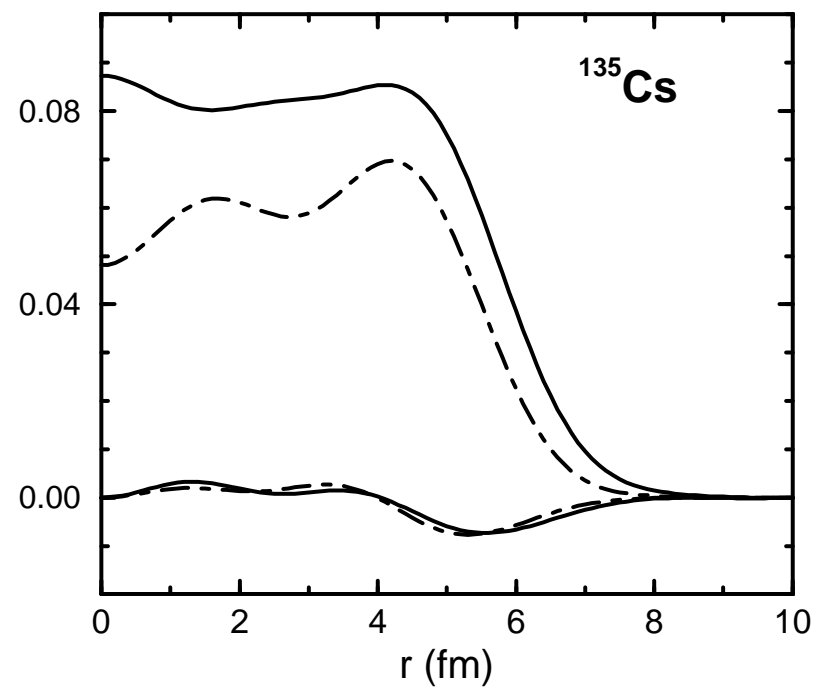
\begin{figure}
\caption{Same as in Fig. \ref{figF}, but for the Cs nuclei.}
\label{figH}
\end{figure}

In Ref.~\cite{Mille} it was suggested that the wide range
of stable isotopes of Yb might be used for a comparison
of PNC observables. These nuclei are generally deformed
and therefore, in order to represent graphically the 
deformed densities, the density functions are 
decomposed in multipole moments
\begin{equation}
\rho({\bf r}) = {1\over \sqrt{4 \pi}}\sum_L \rho_L(r) 
Y_{L0}(\theta, \phi).
\end{equation}
The monopole $L=0$, quadrupole $L=2$, and hexadecupole $L=4$
components of the neutron density in $^{174}$Yb are shown
in Fig.~\ref{figE}. The deformations correspond to the 
neutron quadrupole moment $Q_n = 11.02~b$, and the neutron
hexadecupole moment $H_n = 0.067~b^2$. Since the 
hexadecupole deformations are generally small, in the
following figures we compare only the monopole and
quadrupole components of the ground-state proton and 
neutron densities. These are shown in Fig.~\ref{figF}
for the even-N Yb isotopes. The deformations generally 
decrease with increasing the neutron number (see the 
$\beta_2$ values in Fig.~\ref{figC}). The difference
between the quadrupole components of the proton 
and neutron density is small for all Yb nuclei in 
the sequence $N=98-106$. The proton and neutron densities
differ much more in the spherical part of the distribution, 
not only on the surface (different radii), but 
especially in the bulk region. Similar features are 
observed for the Ba (Fig.~\ref{figG}) and Cs (Fig.~\ref{figH})
isotopes. These nuclei are less deformed, and  
the spherical approximation for the densities will work
better than in the case of the Yb isotopes. Correspondingly,
the differences in the quadrupole components of the 
neutron and proton densities are smaller. The monopole
components of these densities,
however, display very different behavior both in the bulk
and the surface region. 

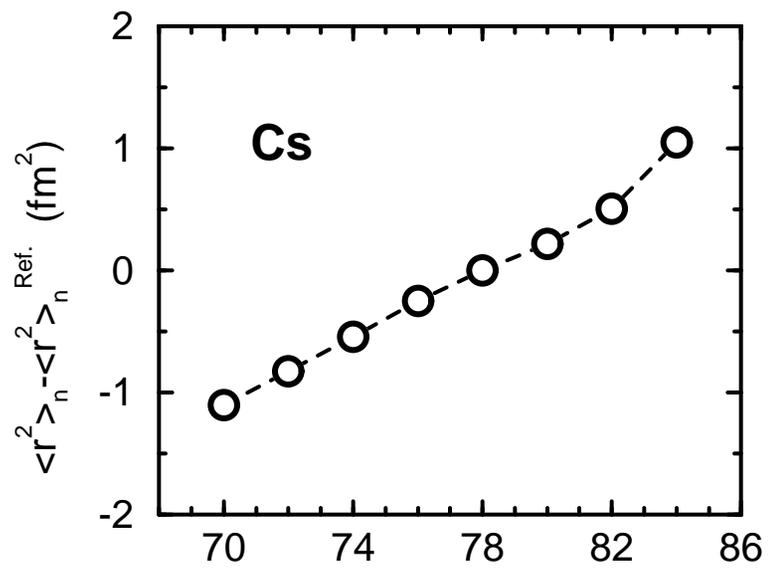
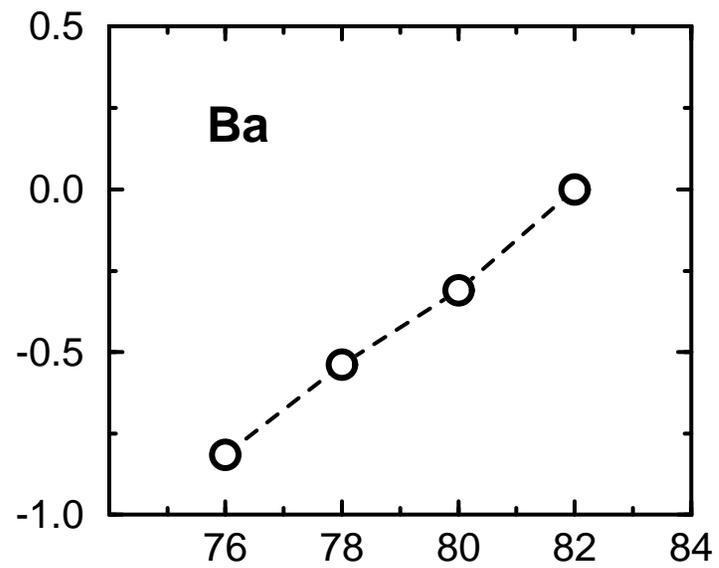
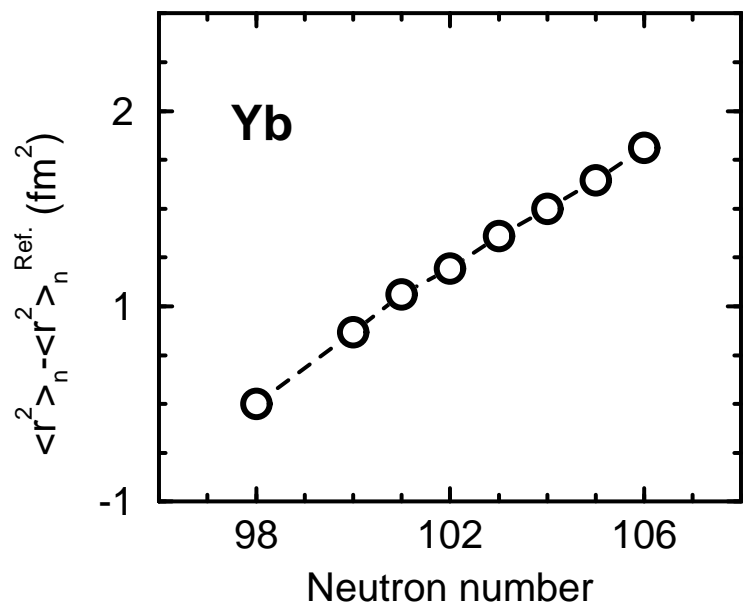
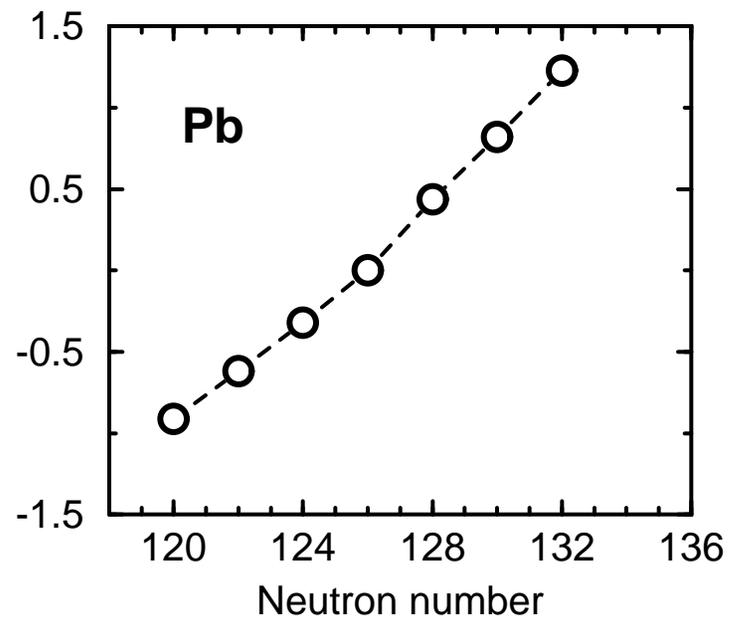
\begin{figure}
\caption{Isotope shifts for the neutron radii of 
the ground-state density distributions of 
Cs, Ba, Yb and Pb nuclei.}
\label{figI}
\end{figure}

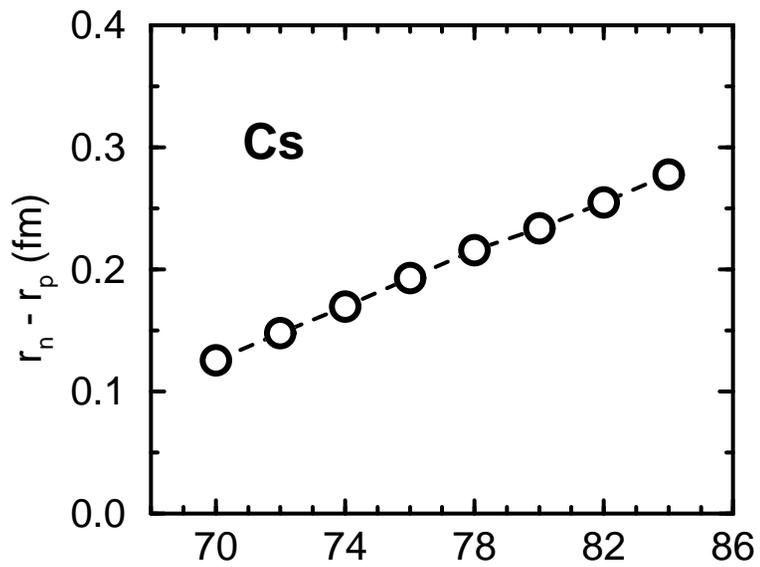
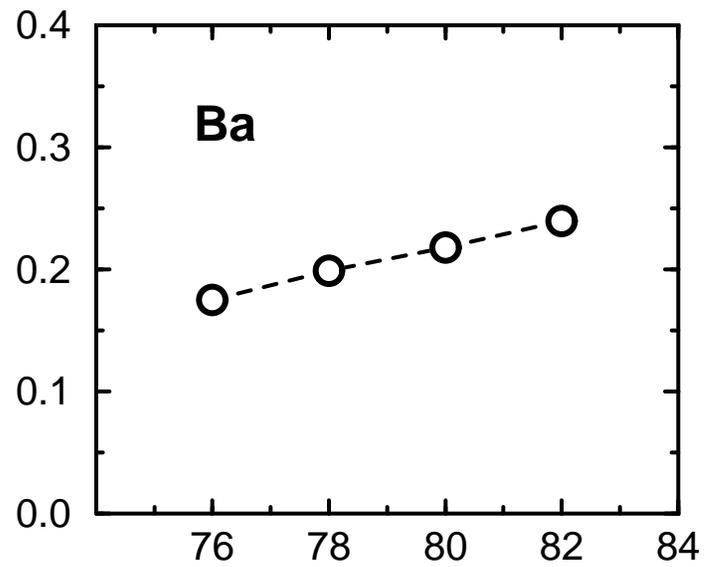
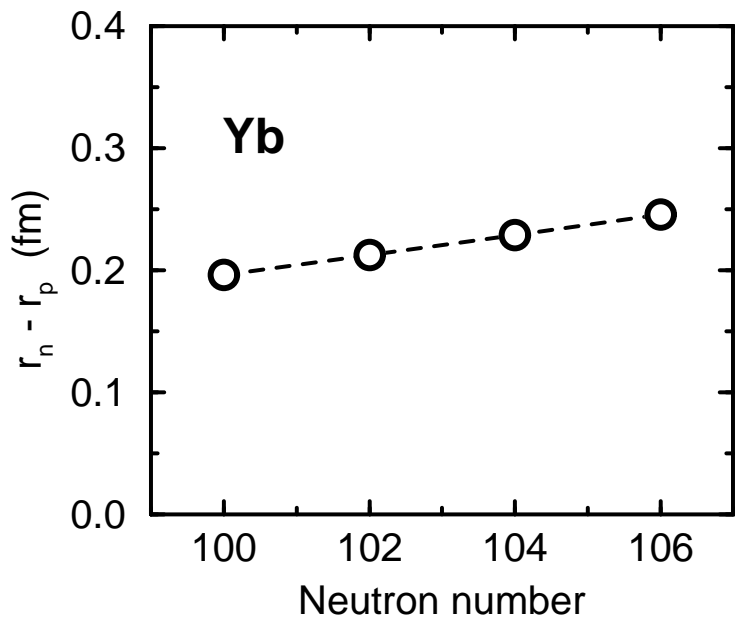
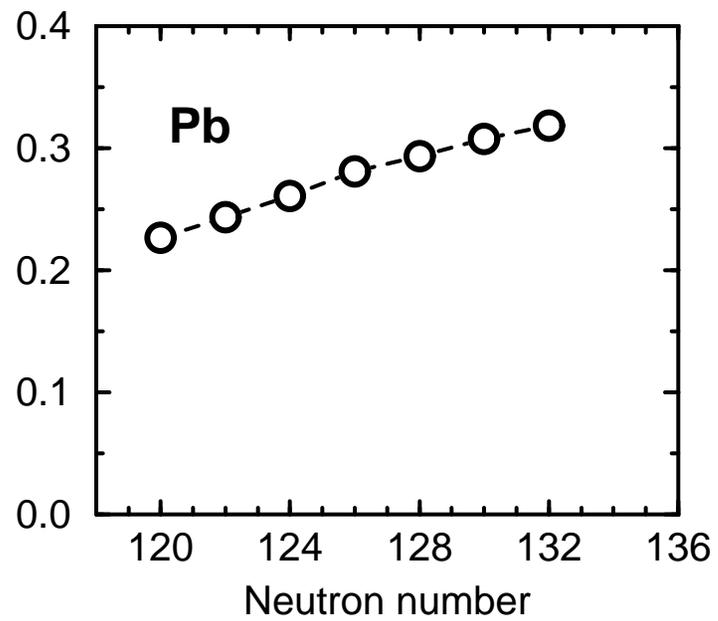
\begin{figure}
\caption{Differences between neutron and proton radii of 
the ground-state density distributions for the
Cs, Ba, Yb and Pb isotope chains.}
\label{figJ}
\end{figure}

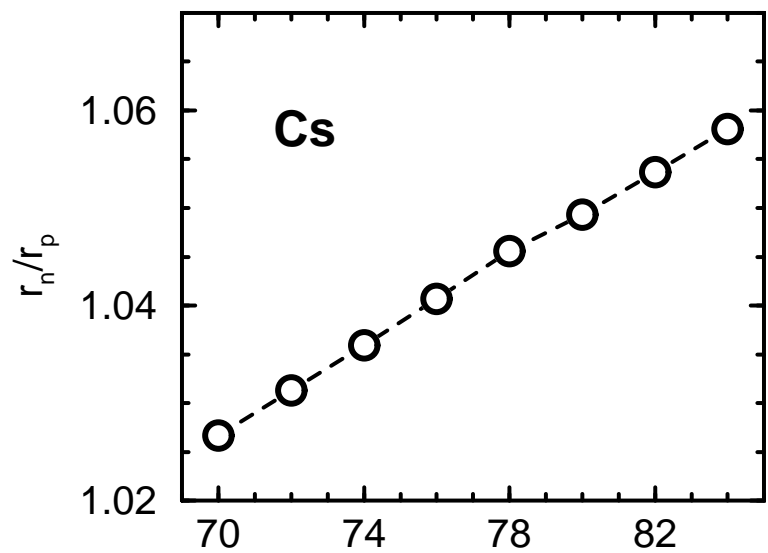
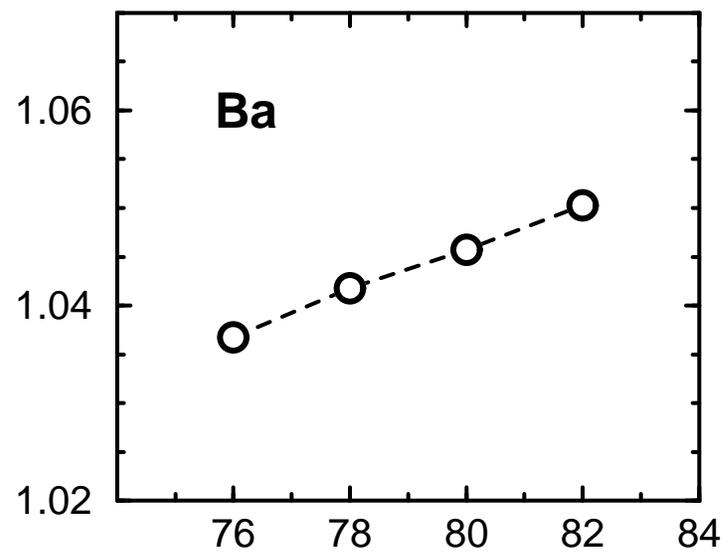
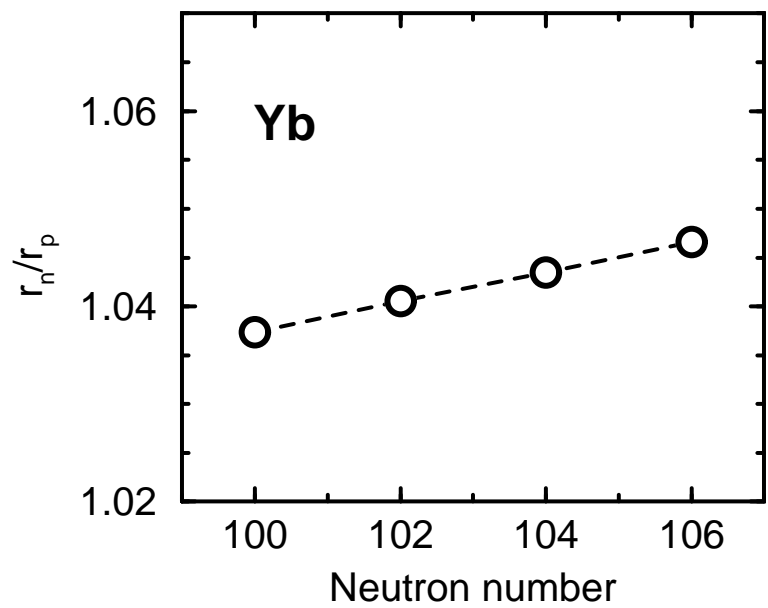
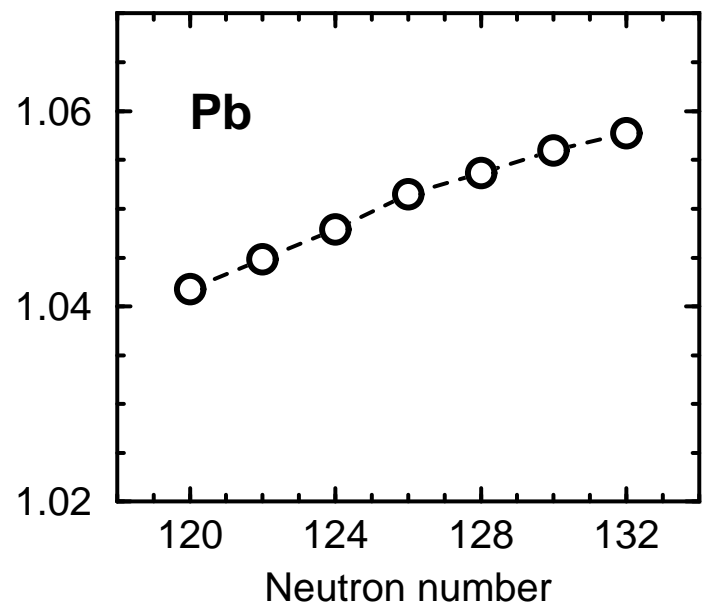
\begin{figure}
\caption{Ratios of neutron to proton radii for 
the ground-state density distributions of the
Cs, Ba, Yb and Pb isotope chains.}
\label{figK}
\end{figure}

The isotope shifts for the neutron radii of 
the ground-state density distributions of Cs, Ba, Yb and Pb nuclei
are shown in Fig.~\ref{figI}. In all four cases 
a uniform increase of the neutron radii is observed. 
The differences between neutron and proton rms radii
are plotted in Fig.~\ref{figJ}, as function of the 
number of neutrons. The differences are of the order
of $0.2 - 0.3$ fm and, of course, increase with the 
neutron number. The increase, however, is not very
steep. For the Yb nuclei, for example, $r_n - r_p$ increases
by less than 0.05 fm between $N=100$ and $N=106$.
The ratios ${r_n/r_p}$ are compared 
in Fig.~\ref{figK}. We notice that while the isotope trend
is similar, the actual values of ${r_n/r_p}$ calculated
with the RHB NL3+D1S model are considerably larger than
those obtained in the Hartre-Fock Skyrme calculations 
for Pb~\cite{Pol.92} and Cs~\cite{PW.99,Chen}. In particular,
for $^{133}$Cs we calculate ${r_n/r_p} = 1.046$, as compared
to the Hartre-Fock values: 1.022 for SkM$^*$, and 1.016 for SkIII.
An approximate estimate for the nuclear
structure correction~(\ref{deltaQ}) to the weak charge has been 
derived in Ref.~\cite{PW.99} 
\begin{equation}
\Delta Q^{n-p}_W \approx  N (Z\alpha)^2 (.221
\epsilon)/q_{{p}}~\,
\end{equation}
where the small parameter $\epsilon$ is defined by
\begin{equation}
(r_{{n}}^2/ r_{{p}}^2) \equiv 1+\epsilon.
\end{equation}
The RHB NL3 value for ${r_n/r_p}$ implies, therefore, 
that the nuclear structure correction to the weak charge
can be more than a factor 2 larger than the estimate
given in ~\cite{PW.99}, bringing the value of $\Delta Q^{n-p}_W$
close to the uncertainties of atomic theory calculations.
On the other hand, the results shown in
Figs.~\ref{figJ} and \ref{figK}  
indicate that the changes in the neutron
distribution parameters along the isotope chains
should not present a limitation on the accuracy with
which the Standard model can be tested in measurements
of isotope ratios of PNC observables.
\bigskip
\begin{center}
{\bf ACKNOWLEDGMENTS}
\end{center}

This work has been supported in part by the
Bundesministerium f\"ur Bildung und Forschung under
project 06 TM 979, by the Deutsche Forschungsgemeinschaft,
and by the Gesellschaft f\" ur Schwerionenforschung (GSI) Darmstadt.
We thank Andreas Wandelt for the relativistic RPA 
calculation of isovector monopole and dipole 
strength distributions in $^{208}$Pb.

\newpage
{\bf Figure Captions}
\bigskip

\end{document}